\documentclass{article}
\usepackage{graphics}
\usepackage{epsfig}
\usepackage{epstopdf}
\usepackage{amsmath}
\usepackage{float}
\begin{document}
\begin{center}
{\Large{ Interacting holographic Ricci dark energy as running vacuum}} \\[0.2in]
Paxy George, Mohammed Shareef.V and Titus K Mathew \\
Department of Physics, \\
Cochin University of Science and Technology, Kochi-22, India.
\end{center}

\abstract
Earlier studies have shown that, in a two component model of the universe with dark matter and the running vacuum energy which is phenomenologically
a combination of $H^2$ and $\dot{H},$ produces either eternal 
deceleration or acceleration in the absence of a bare constant in the density of the running vacuum. In this paper we have shown that, 
in the interaction scenario, where 
the interaction between matter and vacuum is introduced through a phenomenological term, the two component model is capable of causing a transition from 
a prior decelerated to a later accelerated epoch without a bare constant in the running vacuum density. On contrasting the model with the cosmological data, we have found that the interaction coupling 
constant, is small enough, for a slow decay of the running vacuum. The model is subjected to dynamical system analysis which revealed that the end 
de Sitter phase of the model is a stable one. We did an analysis on the thermal behavior of the system, which shows that the entropy is bounded at 
the end stage so that 
the system is behaving like an ordinary macroscopic system. Apart from these we have also performed the state finder diagnostic analysis which implies the 
quintessence nature of running vacuum and confirms that the model will approach the standard $\Lambda$CDM in the future

\section{Introduction}
              One of the remarkable observational findings of modern cosmology is that
the present universe is undergoing an accelerated
expansion\cite{riess,per,sper1,sper2,teg,selj}. The acceleration might caused
by a new
component called dark energy(DE), the nature of which is still not clear. The
simplest model to describe DE is the cosmological constant $\Lambda$, which is the essential ingredient of the $\Lambda$CDM,
the concordance model. But this model is plagued with two major issues, the cosmological constant problem and
coincidence problem. The first problem is refers to the huge discrepancy in the magnitude of $\Lambda$ existing 
between the theoretical prediction from quantum field theory 
and the observational values.

The second one is about the
coincidence of the densities of dark matter and dark energy during the current epoch,
irrespective of their different evolutionary behaviors. There are no explanations for both of these problems in the $\Lambda$CDM model. 
This motivates the dynamical dark energy models, which search other sources of DE beyond the cosmological constant.
The major classes in these types of theories are the quintessence\cite{Fujii}, in which the equation of state can go below the limit

In recent years, models with time dependent vacuum energy density, $\Lambda(t),$ which was named as running 
vacuum energy (RVE)\cite{shapiro1}, 
is become promising, since it  
could have a high enough value in the early stage of the universe to drive the inflation and decays as the universe expands, 
to a small value as observed 
today, which can then cause the late acceleration. The RVE can be extracted using the renormalization group(RG) methods in 
quantum field theory(QFT) in curved space-time\cite{nelson,buch,toms}. A simple Lagrangian description of these 
models at the fundamental level of scalar fields is not yet available, however attempts have been initiated for this
\cite{sola2,sola2013}. In the phenomenological level, the structure of RVE, in the context of the late acceleration, 
can consists of a combination of
$H^2  \, \textrm{and} \, \dot H  $ along with a bare cosmological constant $c_0$\cite{Valent3}. The bare cosmological 
constant is so significant, that the transition into the late accelerated epoch is not possible without it's presence in the energy 
density\cite{Basilakos3}. In refrence \cite{Sola4}, the authors have argued that the recent data, strongly prefer the RVE cosmology over 
the standard $\Lambda$CDM with regards to the late evolution of the universe.

Another interesting approach to explain the late acceleration of the universe, is the holographic Ricci dark energy (HRDE),
inspired by the application of the fundamental holographic principle to the universe as a whole, such that the energy density of the 
respective cosmic component is inversely proportional to the square of an appropriate length scale characterizing the universe. An early study of holographic dark energy has been found in paper \cite{nojiri}.This model has a 
strong analogue with the RVE, in the phenomenological structure that its energy density is also a combination of $H^2 \, \textrm{and} \, \dot H.$ Plenty of studies have 
made on HRDE, by taking the coresponding equation of state as a varying quantity\cite{cohen,xu1,li2,hsu1,gao1,praseetha1}.

A comparison of the RVE and HRDE, where HRDE is also considered as a running energy density, is discussed in reference\cite{mariam}. The authors have claimed that, unlike 
RVE model, HRDE running energy model, leads to unwanted big-rip epoch if the bare cosmological constant is positive and it will leads to the conventional asymptotic 
de Sitter epoch only if the cosmological constant is negative. However they arrived at this conclusion by assuming that each component  
in the model, dark energy and dark matter, are 
separately conserved or self conserved. In a previous work\cite{paxy} we have shown that, if one assume common 
conservation law, satisfied by all the components together, then running HRDE model can 
leads to a conventional evolutionary status for the universe with a prior decelerated epoch followed by an asymptotic de Sitter epoch. In this case also a bare cosmological 
constant is essential for the transition from a prior decelerated epoch to a later accelerated phase.

In the present work, by introducing specific form for the interaction between the dark sectors of the energy density components, in accordance with the 
total conservation of energy, we have shown that, running HRDE model can show a conventional evolutionary behavior, i.e. having a transition to
the late accelerating universe, even without 
the presence of a bare cosmological constant in the energy density.
For simplicity we have chosen the interaction term
as, $Q=bH\rho_{m},$ where $b$ is the interaction parameter. 
The model predicts an end de Sitter phase. We
also perform a dynamical system analysis of the model with a
suitably constructed phase-space and found that the late
acceleration phase is asymptotically stable. This was further substantiated with the study on the evolution of the entropy rate.
This paper is organized as follows.

\section{Running interacting HRDE model }

Holographic dark energy is based on the holographic
principle\cite{bousso,maldacena} which is an important result of quantum
gravity\cite{hooft}. Cohen et.al\cite{cohen} have proposed that
the total energy in a region of size L should not exceed the mass of
a black hole of same size i.e., $L^{3}\rho_{\Lambda}\leq
LM_{p}^{2}$, where $M_{p}^{-2}=8\pi G$ is the reduced Planck mass and $\rho_{\Lambda}$ is the quantum zero point 
energy caused by the UV cut off. In the case of the universe, the cut off length is chosen as IR cut off, which saturate 
this inequality and $\rho_{\Lambda}$ then become the cosmological dark energy, as,
\begin{equation}
\rho_{\Lambda}=\frac{3c^{2}M_{p}^{2}}{L^{2}},
\end{equation}
where $c$ is a constant whose value can be determined by
observations\cite{li1} and $\rho_{\Lambda}$ is said to be the holographic dark energy
density\cite{cohen,li2}.
 This relation clearly indicate that there exist a duality between UV cut
off and IR cut off which, in turns relate the vacuum energy with the 
length scale of the universe\cite{xu1}.
There exist different choices for the IR cut off in the recent 
literature, like the Hubble horizon, particle horizon and future horizon. On taking either Hubble horizon or particle horizon as length scales, it turned out that the corresponding model does not predict late acceleration of the universe\cite{hsu1}. On the other hand, 
if one take the future event horizon as the cut off scale, then the corresponding model will suffer from the problem of causality violation\cite{li1,feng}.   
Later Gao
et al suggested that Ricci scalar can be taken as a safe IR cut off, where the corresponding model will be free from the above two problems\cite{gao1}. This dark energy given by 
the relation  
\begin{equation}\label{equ:3}
\rho_{hrde}(H)=3M_P^2\beta (2H^{2}+\dot{H}),
\end{equation}
and is dubbed
as the holographic Ricci dark energy($hrde$), where $H$ is the Hubble parameter, $\beta$ is a dimensionless parameter 
characterizing the running of the energy density 
and the over-dot represents a derivative with respect to cosmic time. 
Comapred to this, the RVE density given in \cite{sola5}, as,
\begin{equation}
 \rho_{rve}(H)=3M_P^2\left(c_0+\nu H^2 + \frac{2}{3}\alpha \dot H\right),
\end{equation}
consists of an additive constant $c_0$ and two arbitrary parameters, 
$\nu \, \textrm{and} \, \alpha.$ But to alleviate 
the general difference between them, one can phenomenologically add a bare constant to the HRDE 
density in equation (\ref{equ:3}) \cite{paxy,mariam}, and may also assume the relations $\nu=\frac{4\alpha}{3}$ and $\beta=\frac{2\alpha}{3}=\frac{\nu}{2}.$
From quantum field theoretic consideration, it has been shown that the running parameters, both $\nu$ and $\alpha,$ are of the 
order $10^{-3}$\cite{sola2}.

 For a spatially flat Friedmann universe, we have the basic evolution equations as,
\begin{equation}\label{equ:4}
3H^{2}=\rho_{m}+\rho_{de},
\end{equation}
where $\rho_{m}$ is the dark matter density and $\rho_{de}$ is dark energy density. We have
 the conservation equations as,
\begin{equation}\label{equ:5}
\begin{split}
\dot{\rho}_{de}+3H(\rho_{de}+P_{de})=-Q,\\
\dot{\rho}_{m}+3H(\rho_{m}+P_{m})=Q,
\end{split}
\end{equation}
where, $Q$ is the interaction term which gives the rate of energy
exchange between dark matter and dark energy. For $Q>0$ the
energy is being transferred from dark energy to dark matter, whereas for
$Q<0$ the transfer of energy is from dark matter to dark energy\cite{pereira}. We consider a specific form for the interaction term, 
$Q=3bH\rho_{m},$ where $b$ is the coupling constant. 
It should be noted that, switching of the phenomenological interaction, by taking $b=0,$ 
is equivalent to the fact that each components, that is dark energy and the non-relativistic matter, are self conserved. 

For running HRDE, the equation of state is $p_{hrde}=-\rho_{hrde}$ and for non-relativistic matter, 
$p_m=0.$ Then from the conservation law (\ref{equ:5}), we have,
\begin{equation}\label{equ:6}
\begin{split}
\frac{d\Omega_{hrde}}{dx}=-3b\Omega_{m},\\
\frac{d\Omega_m}{dx}=-3(1-b) \Omega_{m},
\end{split}
\end{equation}
 where we have changed the variable from $t$ to $x=ln a.$
The above equation implies that, $\Omega_m=\Omega_{m0}e^{-3(1-b)x},$ where $\Omega_{m0}=\rho_{mo}/3H_0^2$ is the present mass density parameter of the non-relativistic cosmic component.

Combining this with equations (\ref{equ:3}) (\ref{equ:4}) and (\ref{equ:6}), it follows,

\begin{equation}\label{equ:10}
\frac{d^{2}h^{2}}{dx^{2}}+3\frac{dh^{2}}{dx}+9b\Omega_{m0}e^{-3(1-b)x}=0,
\end{equation}
where $h=H/H_0,$ $H_0$ is the present value of the Hubble parameter.
 The solution of the
above second order differential equation gives the evolution of the weighted Hubble parameter as,
\begin{equation}\label{equ:11}
h^{2}=\frac{\Omega_{m0}}{1-b} e^{-3(1-b)x} -\frac{1}{3}C_1e^{-3x}+C_2.
\end{equation}
The constant coefficients $C_{1}$ and $C_{2}$ can be determined by the initial
conditions,
\begin{equation}\label{equ:12}
h^{2}\mid_{x=0}=1, \hspace{0.3in}
\left.\frac{dh^{2}}{dx}\right|_{x=0}=\frac{2\Omega_{hrde0}}{\beta}-4,
\end{equation}
where  $\Omega_{hrde0}$ is the current mass density parameter
 corresponding to running HRDE.
The values of the coefficients are then found to be,
\begin{equation}\label{equ:13}
\begin{split}
C_{1}=\frac{2\Omega_{hrde0}}{\beta}+3\Omega{m0}-4,\\
C_{2}=\frac{2\Omega_{hrde0}}{3\beta}-\frac{b}{1-b}\Omega{m0}-\frac{1}{3}.
\end{split}
\end{equation}
In the limit $x\to -\infty \, (\textrm{equivalently} \, a \to 0),$ the Hubble parameter in equation (\ref{equ:11}) behave as, 
$h^2 \to \left(\frac{\Omega_{m0}}{1-b} a^{3b} -\frac{1}{3} C_1\right)a^{-3},$ where the coefficient, in principle, depends on the scale factor, however 
it implies a decelerated expansion.
But the interaction parameter, $b<<1$ for a slowly decaying vacuum (in fact that is the case - see parameter extraction section)  $a^b \sim 1$, so the coefficient is approximately a constant 
$\left(\frac{\Omega_{m0}}{1-b} -\frac{1}{3} C_1\right),$ then also the Hubble parameter 
represent a matter dominated decelerated universe in the above limit.
In the future, limit $x\to +\infty, \, (\textrm{equivalently} \, a\to\infty),$ the Hubble parameter tends to constant, $h^2\to C_2$ corresponds to de-Sitter phase. 
This shows that the model predicts a transition from an early decelerated epoch to a later accelerated epoch in the 
expansion history of the universe. 

On the other hand, if one choose $b=0$ then equations (\ref{equ:5}), reduces to independent conservation laws for  
dark energy and dark matter consequently we have constant vacuum density and varying matter density. The situation is then similar to the standard $\Lambda$CDM model.

On taking $b=0$ in the Hubble parameter in equation(\ref{equ:11}) we get,

\begin{equation}
 h^2=\left(\Omega_{m0} - C_1/3\right) e^{-3x} + C_2=\tilde{ \Omega}_{m0} e^{-3x} + C_2,
\end{equation}
where\begin{equation}
\tilde{\Omega}_{m0}=\frac{4}{3}\left(1-\frac{\Omega_{hrde0}}{2\beta} \right),      
 \end{equation}
  is the effective mass parameter (such kind of effective mass parameter can be seen in references \cite{sahni1,starobinsky}) 
  corresponds to the non-relativistic matter. The above Hubble parameter equation mimics the corresponding $\Lambda$CDM equation with a cosmological constant $C_2$ and present mass density parameter $\tilde{\Omega}_{m0}$ and it does implies an early deceleration and a late 
  acceleration.

\section{Parameter Estimation} We estimate the best fit of the
model parameters $\beta,b$ and $H_0$ using the combined data set, consisting of the supernova data set, the Cosmic Microwave
Background (CMB) data from the WMAP 7-yr and Planck 2013 observations and the Baryon Acoustic Oscillation
(BAO) data from Sloan Digital Sky Survey(SDSS). Type Ia supernova observations, 
compiled in\cite{kowaslski}, composed of 13 independent data sets, a total of 307 data points showing the magnitudes of
supernovae at different red shifts. The method involved is the comparison of the observed distance modulus with the 
theoretical value predicted from the model. We have the distance modulus as a function of the model parameters,
\begin{equation}\label{equ:15}
\mu_{t}(\beta,b,H_0,z_{i})=m-M\\=5\log_{10}\left[\frac{d_{L}(\beta,b,H_0,z_i)}{Mpc}\right]+25,
\end{equation}
where $m$ and $M$ are the apparent and absolute magnitudes of the supernovae(SNe) respectively, $z_i$ is the red shift of the
\begin{table}[ph]
\renewcommand{\arraystretch}{2.3}
\caption{parameter estimation using WMAP7 and Planck data.}\centering
\begin{tabular}{|c |c |c| }
\hline
$parameters$ & $SNe+CMB(WMAP7)+BAO(SDSS)$ & $SNe+CMB(Planck)+BAO(SDSS)$ \\ 
\hline
$\beta$  &  0.4622 & 0.4612 \\
$b$ &  0.0076& 0.0094 \\
$H_0$ & 70.06 & 70.18 \\ 
\hline
\end{tabular} \label{table:2}
\end{table}

 supernova and $d_{L}$ is the luminosity distance, defined as,
\begin{equation}\label{equ:14}
d_{L}(\beta,b,H_0,z_i)=c(1+z_i)\int_{0}^{z_i}\frac{dz}{H(\beta,b,H_0,z)},
\end{equation}
where $H(\beta,b,H_0,z)$ is the Hubble parameter as a function of the model parameters and $c$ is the speed
of light. One can now have the $\chi^{2}$ function which include theoretical and observational magnitude as,
\begin{equation}\label{equ:16}
\chi^{2}(\beta,b,H_0)=\sum_{i=1}^{n}\frac{[\mu_{t}(\beta,b,H_0,z_{i})-\mu_{i}]^{2}}{\sigma_{i}^{2}},
\end{equation}
where $\mu_{i}$ is the observational distance moduli for the
$i^{th}$ Supernova, $\sigma_{i}^{2}$ is the standard variation of the
observation and $n=307,$ the total number of data points. 
This function can then minimized to get the best fit values of the model parameters.

For obtaining the $\chi^2$ function we also used 
Background (CMB) data from the WMAP 7-yr observation and the Baryon Acoustic Oscillation
(BAO) data from Sloan Digital Sky Survey(SDSS)\cite{Wang1}. The BAO signal has been directly
detected by SDSS survey at a scale $\sim$100MPc. The BAO peak parameter value was first proposed by
D. J. Eisenstein, et al\cite{eisenstein} and is defined as
\begin{equation}\label{equ:17}
\mathcal{A}=\frac{\sqrt{\Omega_{m}}}{h(z_{1})^{\frac{1}{3}}}\left(\frac{1}{z_{1}}\int_{0}^{z_{1}}\frac{dz}{h(z)}\right)^{\frac{2}{3}},
\end{equation}
Here  h(z) is the Hubble parameter, $z_{1} = 0.35$ is the red shift of
the SDSS sample\cite{tegmark}. Using SDSS data from luminous red
galaxies survey the value of the parameter $\mathcal{A}$(for flat
universe) is given by $\mathcal{A} = 0.469 \pm 0.017$\cite{eisenstein}. The
$\chi^{2}$ function for the BAO measurement takes the form
\begin{equation}\label{equ:18}
\chi^{2}_{BAO}=\frac{(\mathcal{A}-0.469)^{2}}{(0.017)^{2}}.
\end{equation}
The CMB shift parameter is the first peak of CMB power
spectrum\cite{bond} can be written as
\begin{equation}\label{equ:19}
\mathcal{R}=\sqrt{\Omega_{m}}\int_{0}^{z_{2}}\frac{dz}{h(z)},
\end{equation}
Here $z_{2}$ is the red shift at the last scattering surface. 
\begin{figure}[pb]
\centerline{\psfig{file=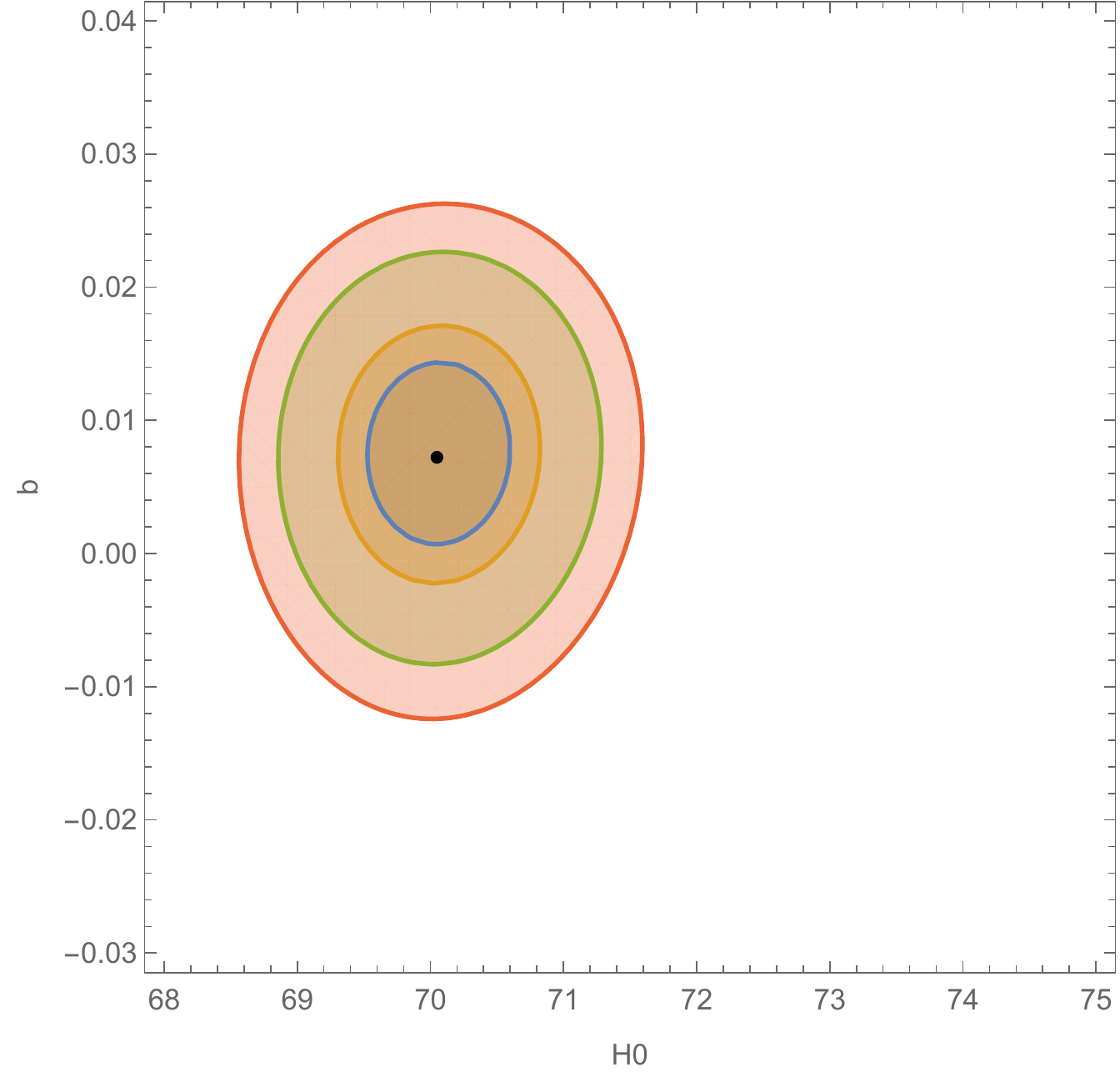,width=4.7cm}}
\vspace*{8pt}
\caption{Confidence intervals for the parameters(b,$H_0$) using the
SNe+BAO+CMB data sets. The point indicate the best estimated values
of the parameters,$b=0.007615_{-0.01}^{+0.01}$,
$H_0=70.06_{-0.52}^{+0.53}$. The confidence intervals shown
corresponds to the 68.3,95.4,99.73 and 99.99 $\%$ of
probabilities.}\label{fig:f3}
\centerline{\psfig{file=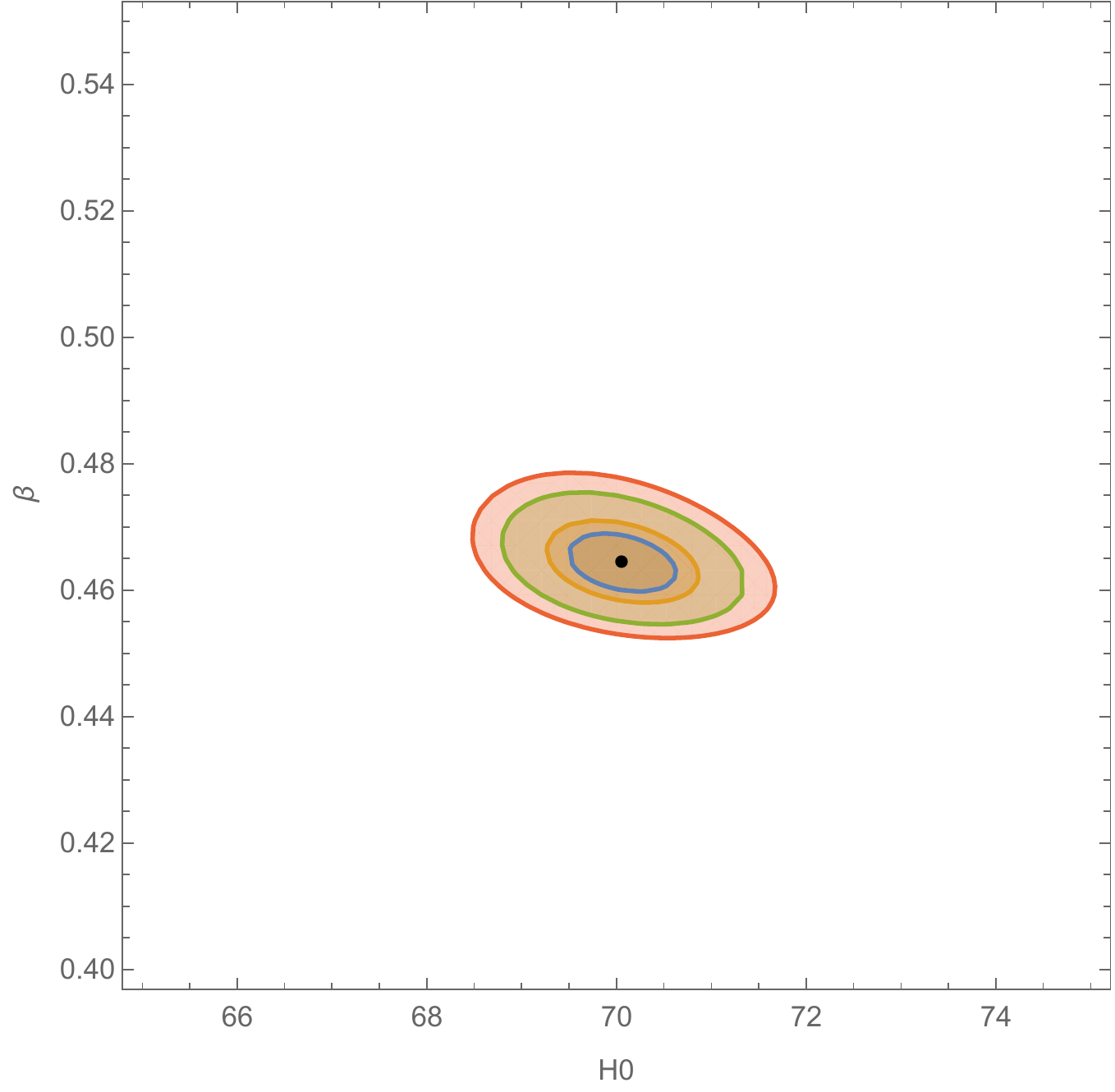,width=4.7cm}}
\vspace*{8pt}
\caption{Confidence intervals for the parameters($\beta,H_0$) using
the SNe+BAO+CMB data sets. The point indicate the best estimated
values of the parameters,$\beta$= $0.46422_{-0.005}^{+0.004}$,
$H_0=70.06_{-0.52}^{+0.53}$. The confidence intervals shown
corresponds to the 68.3,95.4,99.73 and 99.99 $\%$ of probabilities.
}\label{fig:f4}
\end{figure}
From the WMAP 7-year data, $z_{2}=1091.3$. At this red shift
$z_{2}$, the value of shift parameter would be $\mathcal{R}=1.725\pm
0.018$\cite{komatsu}. The $\chi^{2}$ function for the CMB measurement can be
written as
\begin{equation}\label{equ:20}
\chi^{2}_{CMB}=\frac{(\mathcal{R}-1.725)^{2}}{(0.018)^{2}}.
\end{equation}
Considering three cosmological data sets together, i.e. (SNe+BAO+CMB),  the total $\chi^{2}$ function is
then given by
\begin{equation}\label{equ:21}
\chi^{2}_{total}=\chi^{2}_{SNe}+\chi^{2}_{BAO}+\chi^{2}_{CMB}.
\end{equation}
The best for the parameters obtained by
minimizing total $\chi^{2}$ are, $H_0=70.06kms^{-1}Mpc^{-1}, \beta=0.4642, b=0.0076.$ 
The minimum of the per degrees of freedom is found to be $\chi^2_{d.o.f}=1.028.$

We have constructed confidence interval planes $(\beta,H_0)$ for constant $b$ and $(b,H_0)$ for constant $\beta$ and are given in figures
(\ref{fig:f3}) and (\ref{fig:f4}) respectively. The confidence
intervals corresponding to 68.3, 95.4, 99.73 and 99.99 $\%$ of
probabilities respectively has been potted. The corrected values of the parameters corresponding to 68.3$\%$ 
probability are, 
$\beta=0.464_{-0.004}^{+0.005}$, $b=0.00762_{-0.01}^{+0.01}$,
$H_0=70.06_{-0.52}^{+0.53}kms^{-1}Mpc^{-1}$ respectively. 
 
 We repeat the above computation by using new observational data on CMB from Planck 2013\cite{planck1,shafer}, additional BAO data from the latest SDSS observation\cite{blake,sdss} and SNe data. The parameter values then obtained corresponding to the $\chi^2_{min}$ are $\beta \sim 0.4612,$ $b \sim 0.0094$ and $H_0 \sim 70.18kms^{-1}Mpc^{-1}$. In reference \cite{fu}, the model parameters $\beta$ and $b$ are constrained using SNe 557 data along with BAO (SDSS) and CMB (WMAP7) data and have obtained $\beta=0.433$ and $b=0.032$. Our value for $\beta$ is close to this
 but the value of $b$ is slightly high.

\section{Evolution of cosmological parameters}
The Hubble parameter 
 in equation(\ref{equ:11}) shows a decreasing behavior with the scale
 \begin{figure}[pb]
\centerline{\psfig{file=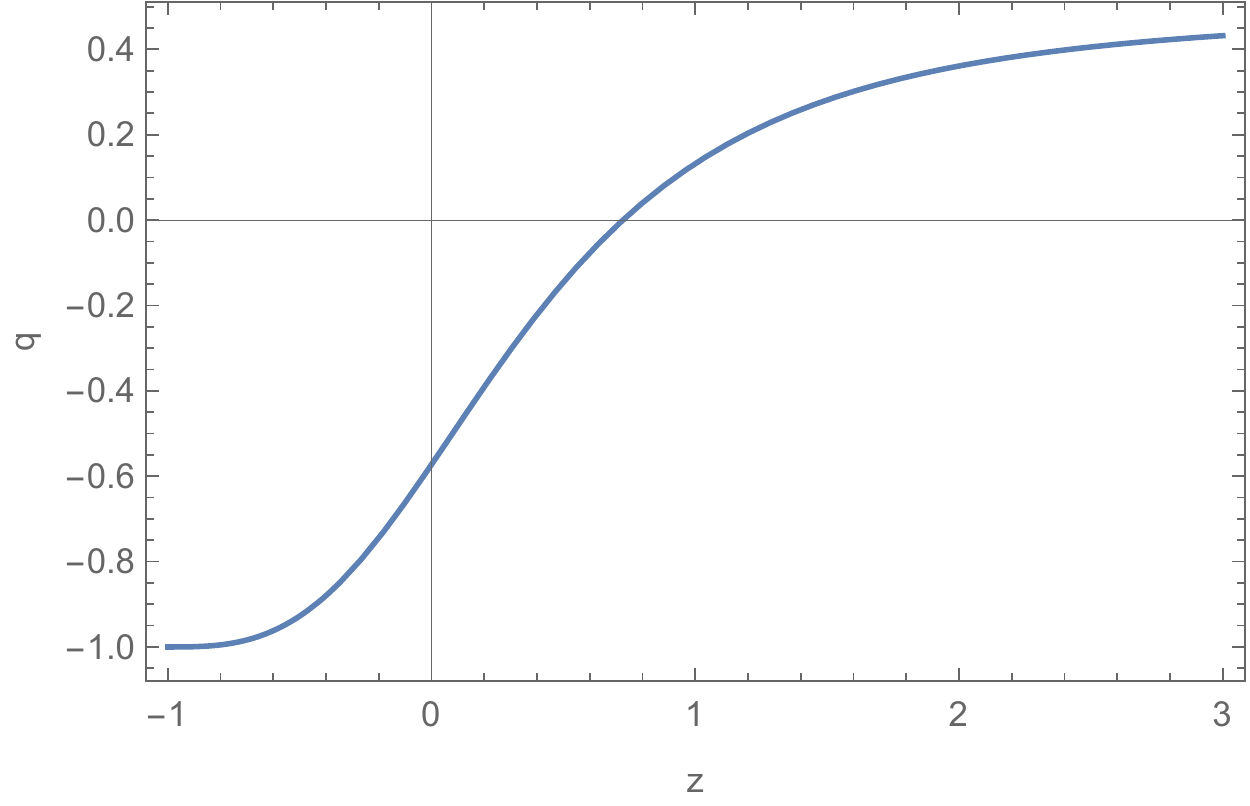,width=4.7cm}}
\vspace*{8pt}
\caption{The evolution of the deceleration parameter with red shift
for SNe+CMB+BAO data. The transition red shift for the best estimated values of the
model parameters is $z_{T}=0.71$.}\label{fig:f6}
\end{figure}
factor. It can have infinitely large value in the early stages and decreases as the universe expands and finally saturated to a 
constant value as $a \to \infty.$ 

 The deceleration parameter which characterizing the decelerating/accelerating nature of the universe can be expressed as,
 \begin{equation}\label{equ:22}
 q=-1-\frac{\dot{H}}{H^{2}}.
 \end{equation}
 Using equation(\ref{equ:11}), the deceleration parameter takes the
 form,
 \begin{equation}\label{equ:23}
 q=-1-\frac{C_1e^{-3x}-3\Omega_{m_0}e^{-3(1-b)x}}{2(\frac{-C_1}{3}e^{-3x}+C_2+\frac{\Omega_{m_0}}{1-b}e^{-3(1-b)x})}.
 \end{equation}
 
 The evolution of the deceleration parameter in accordance with the SNe+CMB+BAO
data sets has been plotted in figures(\ref{fig:f6}).
\begin{figure}[pb]
\centerline{\psfig{file=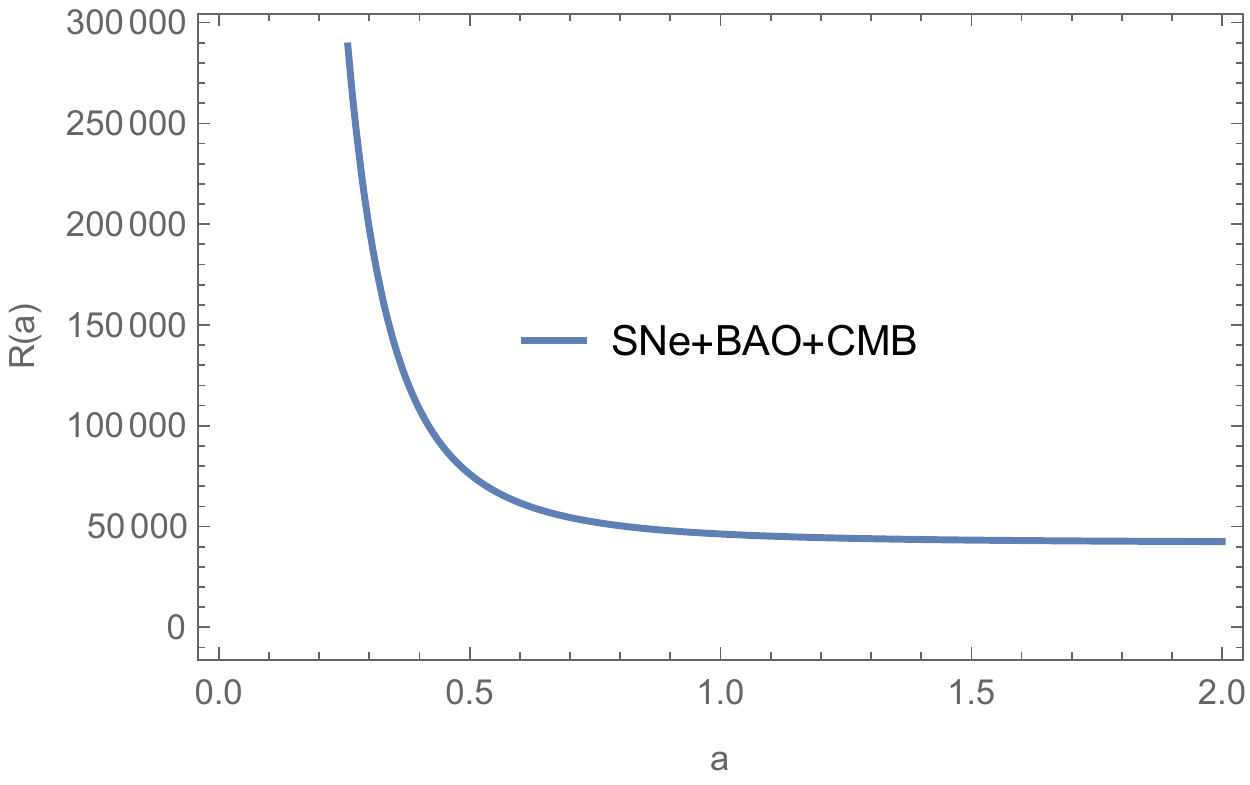,width=4.7cm}}
\vspace*{8pt}
\caption{The evolution of curvature scalar with scale factor for the
best estimate parameters.}\label{fig:f8}
\end{figure}

The transition red shift is found to be $z_T=0.71$ for SNe+CMB(WMAP)+BAO data and is sightly high $z_{T}=0.74$ for SNe+CMB(Planck2013)+BAO data. Both these are comparable with the range of the transition red shift,$z_{T}=0.45-0.73$ in the concordance $\Lambda CDM$ model.\cite{alam}. 

The present value of deceleration parameter
corresponds to $z=0$ is,
\begin{equation}\label{equ:27}
q_{0}=-1-\frac{C_{1}-3\Omega_{m_0}}{2(\frac{C_{1}}{3}+C_{2}+\frac{\Omega_{m_0}}{1-b})}.
\end{equation}
For the best estimated model parameters,  
$q_0=-0.572^{+0.015}_{-0.014}$
for the SNe+CMB(WMAP)+BAO data and $q_0=-0.58$ for SNe+CMB(Planck2013)+BAO data. The corresponding WMAP value is around 
$q_{0}\sim-0.60$\cite{wmap}.

The evolution of the mass density parameter from the conservation equation
of matter(\ref{equ:6}) is,
\begin{equation}\label{equ:29}
\Omega_{m}=\Omega_{m0}a^{-3(1-b)}.
\end{equation}

 Because of the smallness of the parameter $b,$ the evolution of $\Omega_m$ is not very much different from the conventional one. 
 Starting from a infinitely 
 high value in the very early stage, and reduces to zero as $a \to \infty.$

The behavior of the Hubble parameter and the mass density parameter indicates the presence of the big-bang singularity at the beginning, which 
clearly confirmed from the nature of curvature scalar parameter, which is defined as,
\begin{equation}\label{equ:30}
R=6(\dot{H}+2H^{2}).
\end{equation}
Substituting for H and $\dot{H}$ from 
equation(\ref{equ:11}), we get the curvature scalar as,
\begin{equation}\label{equ:31}
R(a)=6H_0^{2}\left[\frac{1}{2}(C_{1}a^{-3}-3\Omega_{m_0}a^{-3(1-b)})+2(\frac{-1}{3}C_1a^{-3}+C_{2}+\frac{\Omega_{m_0}}{1-b}a^{-3(1-b)})\right].
\end{equation}
In the limit 
$a\rightarrow 0,$ the curvature scalar, $R \to \infty$ (at which density as given by equation (\ref{equ:29}) is also tending to infinity) indicating that there was a big-bang in the past. A confirmation regarding the big-bang at the origin should come from the inclusion of the radiation component also, which is beyond the scope of the present analysis. A graphical representation of the evolution of the curvature scalar
is shown in
figure(\ref{fig:f8}).

\section{State Finder Analysis}
The state finder diagnostic method, proposed
by Sahini et al.\cite{sahni} is an effective geometric tool to
compare the evolutionary properties of a given dark energy model in comparison with other models, especially $\Lambda$CDM. 
The state finder parameters, $r$ and $s,$ are defined as,
\begin{figure}[pb]
\centerline{\psfig{file=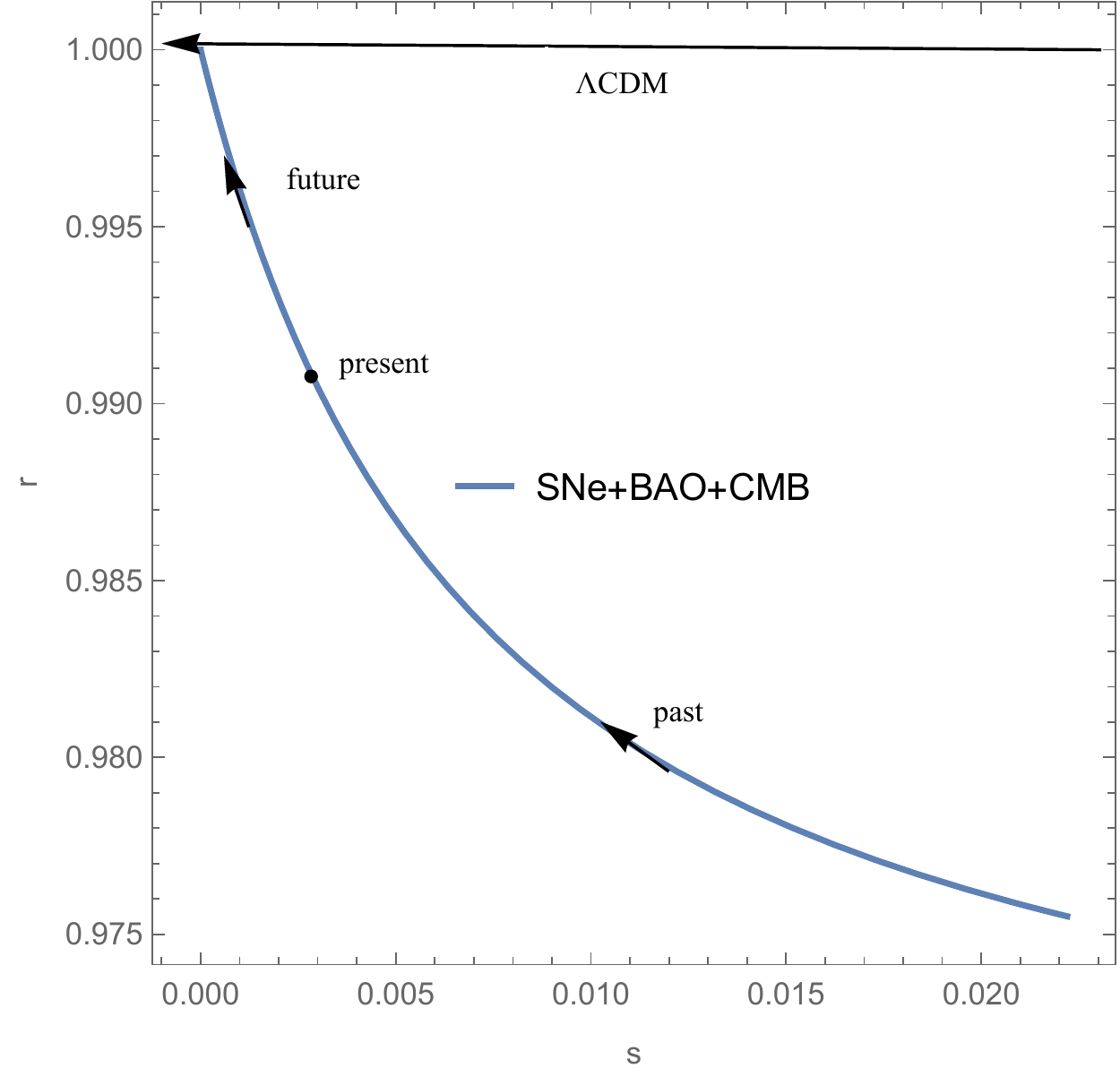,width=4.7cm}}
\vspace*{8pt}
\caption{The plot of $(r,s)$ parameter for the best estimate
parameters for SNe+CMB+BAO data sets.}\label{fig:f10}
\end{figure}
\begin{equation}\label{equ:32}
\begin{split}\\
r=\frac{\dddot{a}}{aH^{3}}=\frac{1}{2h^{2}}\frac{d^{2}h^{2}}{dx^{2}}+\frac{3}{2h^{2}}\frac{dh^{2}}{dx}+1\\
s=\frac{r-1}{3(q-\frac{1}{2})}=-\frac{\frac{1}{2h^{2}}\frac{d^{2}h^{2}}{dx^{2}}+\frac{3}{2h^{2}}\frac{dh^{2}}{dx}}{\frac{3}{2h^{2}}\frac{dh^{2}}{dx}+\frac{9}{2}}.
\end{split}
\end{equation}
For $\Lambda CDM$ model, these parameters constitute a fixed point,
$(r,s) = (1,0)$ in the $r-s$ plane.
 On substituting for the Hubble parameter from  equation(\ref{equ:11}), the parameters for the present model become,
\begin{equation}\label{equ:33}
\begin{split}
r=1+\frac{3(C_1a^{-3}-3\Omega_{m_0} a^{-3(1-b)})}{2(C_2-\frac{1}{3}C_1a^{-3}+\frac{\Omega_{m_0}a^{-3(1-b)}}{1-b})}+\frac{-3C_1a^{-3}+9(1-b)\Omega_{m_0}a^{-3{1-b}}}{2(C_2-\frac{1}{3}C_1a^{-3}+\frac{\Omega_{m_0}a^{-3(1-b)}}{1-b})},\\
s=-\frac{\frac{3(C_1a^{-3}-3\Omega_{m_0} a^{-3(1-b)})}{2(C_2-\frac{1}{3}C_1a^{-3}+\frac{\Omega_{m_0}a^{-3(1-b)}}{1-b})}+\frac{-3C_1a^{-3}+9(1-b)\Omega_{m_0}a^{-3{1-b}}}{2(C_2-\frac{1}{3}C_1a^{-3}+\frac{\Omega_{m_0}a^{-3(1-b)}}{1-b})}}{\frac{9}{2}+\frac{3(C_1a^{-3}-3\Omega_{m_0} a^{-3(1-b)})}{2(C_2-\frac{1}{3}C_1a^{-3}+\frac{\Omega_{m_0}a^{-3(1-b)}}{1-b})}}.
\end{split}
\end{equation}
The above equations shows that, the present model approach $\Lambda$CDM asymptotically, as  $(r,s)\to(1,0),$ when $a\rightarrow\infty.$
The value of the state finder parameters corresponding to the 
present epoch are found to be,
\begin{equation}\label{equ:34}
\begin{split}
 (r_{0},s_{0})_{SNe+CMB(WMAP7)+BAO}=(0.9907,0.00287),\\
 (r_{0},s_{0})_{SNe+CMB(Planck 2013)+BAO}=(0.9886,0.0035).
 \end{split}
\end{equation}
This clearly indicate that the present model is distinguishably
different from the $\Lambda CDM$ model. The evolution of the present model in the $(r,s)$ plane
is shown
in the figures(\ref{fig:f10}).

The trajectory shows that, until it reaches the fixed $\Lambda$CDM point, the parameters 
lies in the region corresponding to $r<1$ and $s>0$ and is resembling the nature of the quintessence dark energy\cite{zhang} and 
opposite to the nature of Chaplygin gas model for which $r>1, \, s<0$\cite{ya}.

\section{Generalized Second Law of Thermodynamics}

\begin{figure}[pb]
{\psfig{file=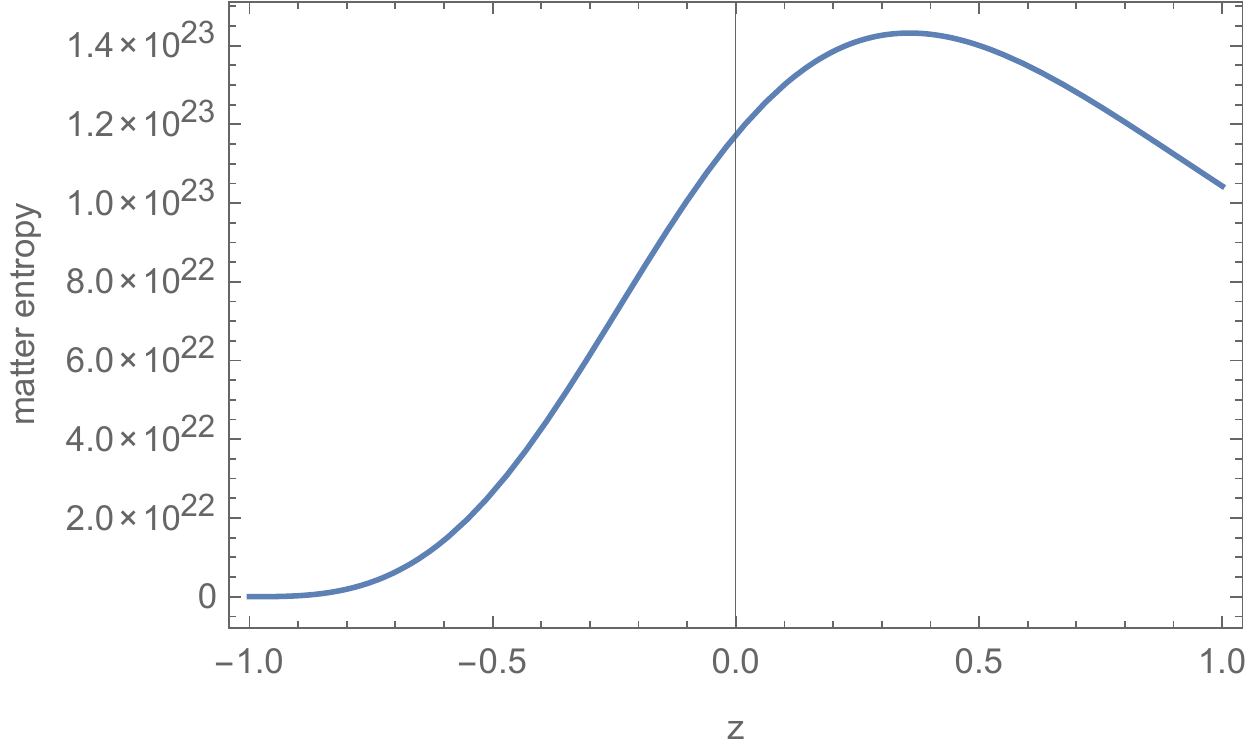,width=4.7cm}} \quad \quad 
{\psfig{file=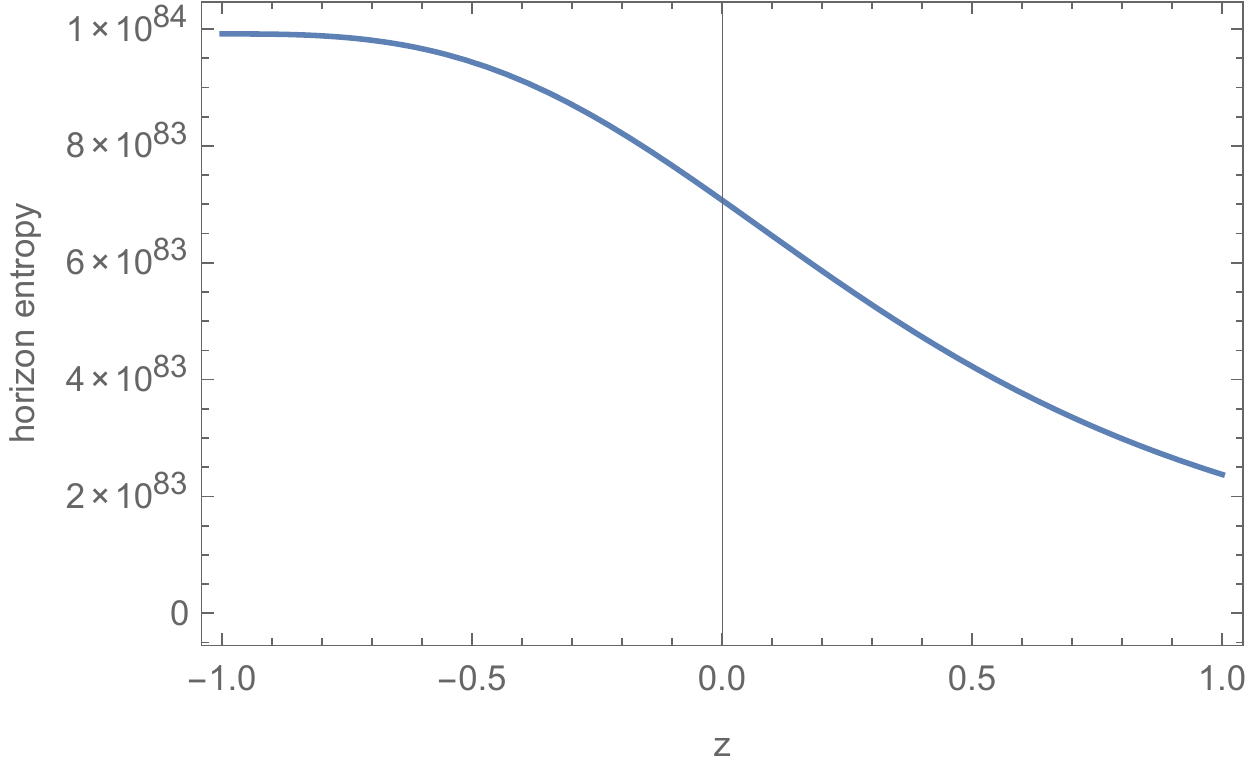,width=4.7cm}}
\vspace*{8pt}
\centering
\caption{Evolution of matter entropy and horizon entropy versus red shift for the best estimate parameters for 
SNe+CMB+BAO data sets.}
\label{fig:f11}
\end{figure}

In this section we are discussing the status of Generalized second
law(GSL) of thermodynamics with 
Hubble horizon  as the thermodynamic boundary. According to GSL, the 
rate change of the entropy of the  horizon plus
entropy of the matter inside the horizon will never
decrease\cite{wang},
\begin{equation}\label{equ:35}
\frac{d}{dt}(S_{H}+S_{m})\geq0,
\end{equation}
where $S_H$ and $S_m$ are the horizon and matter entropy respectively. In general the entropy evolution of a cosmic fluid, is given by,
\begin{equation}
TdS=d(\rho V)+ p dV,
\end{equation}
which must satisfy the integrability condition\cite{maz},
\begin{equation}
\frac{\partial^2S}{\partial T \partial V}=\frac{\partial^2 S}{\partial V \partial T},
\end{equation} 
where $T$ is the temperature and $V$ is the volume bounded by the horizon. From these relations, it can be shown that\cite{Orlando}, 
\begin{equation}
dS=d\left[\frac{(\rho+p)V}{T} + \textrm{constant} \right],
\end{equation}
where ${'constant'}$ appearing from the respective integration. From this it naturally follows that, 
\begin{equation}\label{equ:s12}
S=\frac{(\rho + p)V}{T},
\end{equation}
which gives the entropy of the cosmic fluid apart from a constant.
The entropy of the non-relativistic matter, which is pressureless,  is given by
\begin{equation}\label{equ:36}
S_m=\frac{\rho_m}{T}V,
\end{equation}
where $\rho_m,T$  are matter density and temperature.  The entropy of matter within the Hubble horizon of radius, 
$r=\frac{c}{H},$ volume, $V=\frac{4}{3}\pi r^{3}$ and temperature, given by Gibbon-Hawking relation,
$T=\frac{H}{2\pi}$ is
\begin{equation}\label{equ:37}
S_{m}=\frac{8 \pi^2 H_0 c^3}{H^4} \Omega_{m_0} e^{-3(1-b)x}. 
\end{equation}
 Here we consider the Hubble horizon as the boundary, which is equivalently the apparent horizon for a flat universe.
The plot of matter entropy versus red shift for the best estimated parameters is first increases in the early stage 
and then decreases as shown in figure(\ref{fig:f11})
(left panel).
\noindent Following the relation (\ref{equ:s12}), entropy of dark energy is zero since $p_{de}=-\rho_{de}.$
\begin{figure}[pb]
\centerline{\psfig{file=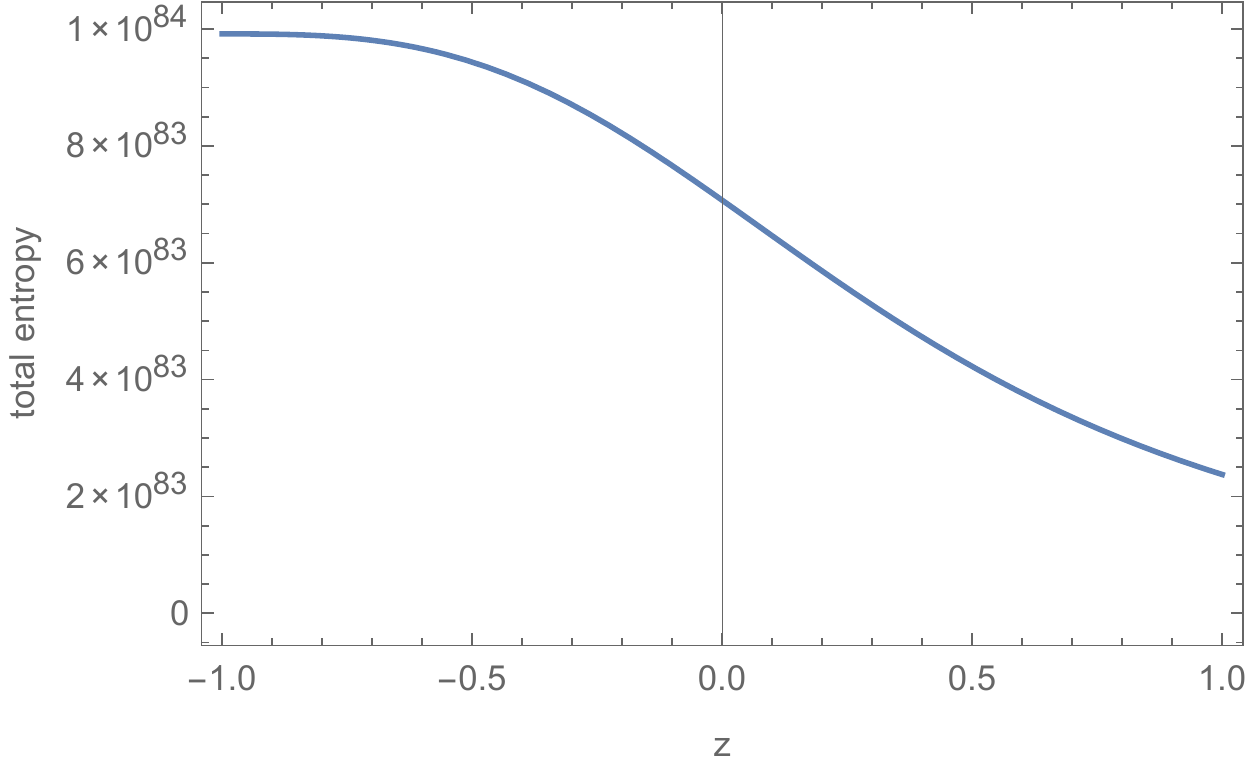,width=4.7cm}}
\vspace*{8pt}
\caption{The plot of validity of generalized second law of
thermodynamics for the best estimate parameters for
SNe+CMB+BAO data sets.}\label{fig:f13}
\end{figure}
The entropy of the horizon is proportional to its area\cite{jacobson}, can be defined as
\begin{equation}\label{equ:38}
S_{h}=\dfrac{A}{4l_p^2},
\end{equation}
where $A=\frac{4\pi c^2}{H^2},$  the area of the event horizon and $l_p$ is the Planck length.  
It then follows as,
\begin{equation}\label{39}
S_{h}=\frac{\pi c^2}{l_p^2 H^2}.
\end{equation}
Using the Hubble parameter in equation(\ref{equ:11}), the evolution of horizon entropy with red shift,
 is given in figure (\ref{fig:f11}) and it always increases.
\begin{figure}[pb]
\centerline{\psfig{file=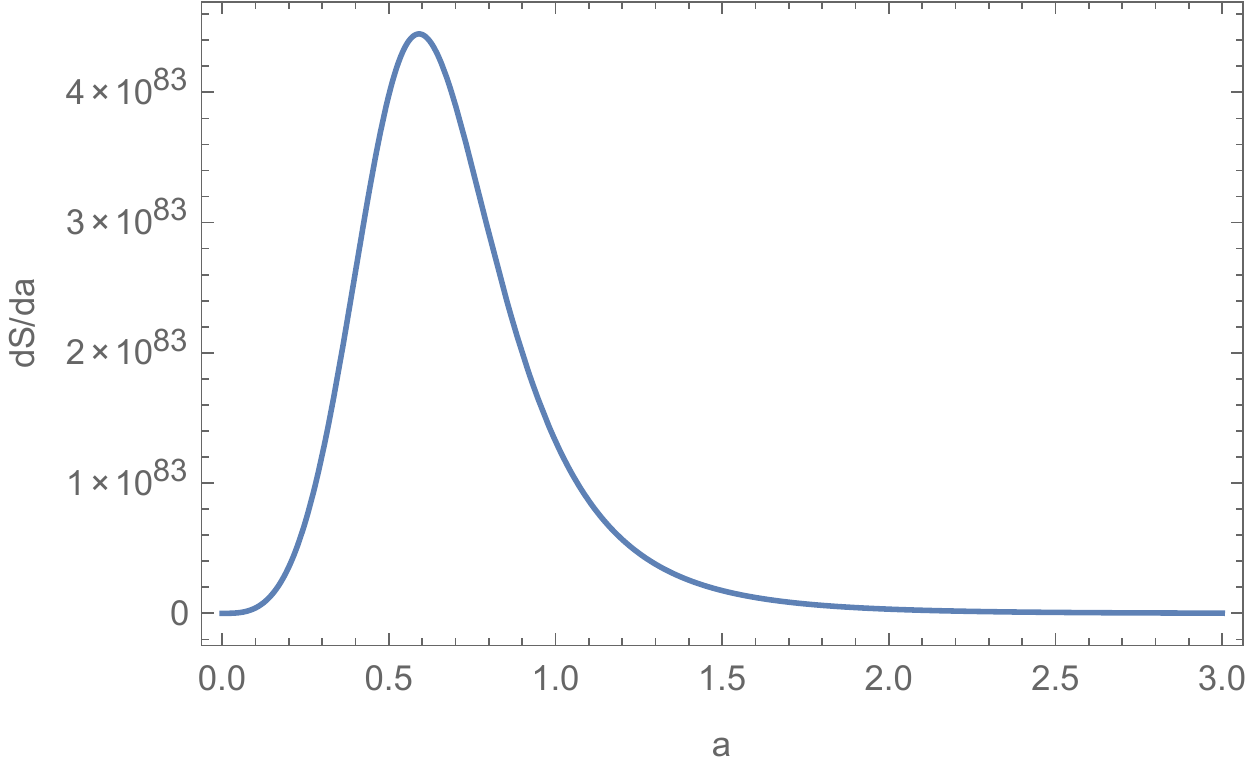,width=4.7cm}}
\vspace*{8pt}
\caption{The plot of rate of entropy change versus scale factor}\label{fig:f14}
\end{figure}
\begin{figure}[pb]
 \centerline{\psfig{file=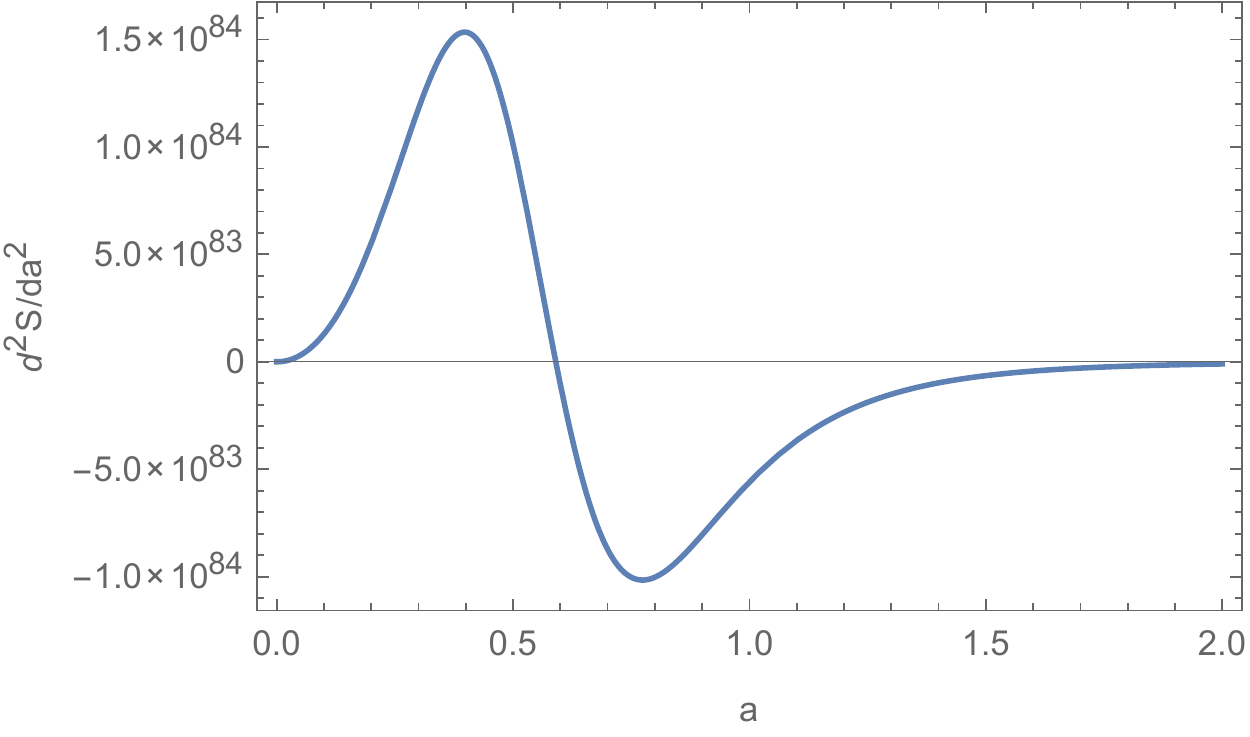,width=4.7cm}}
\vspace*{8pt}
\caption{The evolution of $S^{\prime\prime}$ with scale factor}\label{fig:f9}
\end{figure}
The expression for total entropy is then becomes
\begin{equation}\label{equ:40}
S_{m}+S_{h}=\frac{8\pi^2H_0^2c^3}{H^4}\Omega_{m_0}e^{-3(1-b)x}+\frac{\pi c^2}{l_p^2H^2}.
\end{equation}
We have checked the validity the GSL by numerically evolving the above equation. The plot of total
entropy versus red shift is shown in the figure \ref{fig:f13}.

\noindent The plot shows that the entropy is always increasing and hence GSL is valid.
This can be confirmed by obtaining the behavior of the rate change of entropy is given by
\begin{equation}\label{equ:entro}
\begin{split}
{S^{\prime}}={S_m^{\prime}}+{S_h^{\prime}}, \hspace{2.5in} \\
{S^{\prime}}=-\frac{8\pi^2H_0^2c^3}{H^4}\Omega_{m_0}3(1-b) a^{-3(1-b)}-\frac{dH}{dx}\left(\frac{8\pi^2H_0^2c^3}{H^4}4\Omega_{m_0}a^{-3(1-b)}+\frac{2\pi c^2}{L_p^2 H^2}\right),
\end{split}
\end{equation}
where $'prime'$ denotes derivative with respect to the scale factor $a=e^x.$ In the above equation,  $\frac{dH}{dx}$ is always negative which makes the second term in the above equation positive. Since horizon entropy is too large than matter entropy, the decrease in the matter entropy due to the first term in the above equation will be compensated by the second term as a result  ${S^{\prime}} \geq 0$ always and thus GSL is always valid. In the figure(\ref{fig:f14}), we have shown the behavior of ${S^{\prime}}$ with scale factor, which indicates that ${S^{\prime}}\to0$ as $a\to \infty.$

The evolution characteristics of $S^{\prime}$ indicating that the entropy become maximum as $a\to\infty,$ which in turn implies
that the end state is an equilibrium state. The equilibrium will be stable if it satisfies $S^{\prime\prime}<0$ atleast in the long 
run in the expansion of the universe so that the entropy is bounded. We have obtained $S^{\prime\prime}$ as 
\begin{equation}
\begin{split}
S^{\prime\prime}=\left(\frac{-3(1-3(1-b))(1-b) 8 \pi^2 H_0 c^3 \Omega_{m_0} }{H^4}\right)a^{-2-3(1-b)}+ \left(\frac{24 (1-b) 8 \pi^2 H_0 c^3 \Omega_{m_0}\frac{dH}{da}}{H^5}\right)a^{-1-3(1-b)}+\hspace{2.5in}\\ 8\pi^2 H_0 c^3 \Omega_{m_0}\left(\frac{20 (\frac{dH}{da})^2}{H^6}-\frac{4 \frac{d^2H}{da^2}}{H^5}\right)a^{-3(1-b)}+\frac{\pi c^2}{x^2}\left(\frac{6(\frac{dH}{da})^2}{H^4}-\frac{2\frac{d^2H}{da^2}}{H^3}\right).\hspace{2.5in}
\end{split}
\end{equation}
For the best estimates of the model parameter the evolution of $S^{\prime\prime}$
 with scale factor $a$ is as shown in figure(\ref{fig:f9})
 
The figure shows that, $S^{\prime\prime}>0$ in the early stages, while the model satisfies the entropy maximization condition
$S^{\prime\prime}<0,$  in the later stages of evolution. 
Note that the $S^{\prime\prime}$ approaching  zero from below in the end stage. 
Any system satisfying the condition $S^{\prime}\geq0$ and $S^{\prime\prime}<0$ at the
end stage, is said to be an ordinary macroscopic system \cite{diego}. It can be concluded that the universe explained by the present model 
is evolved like an ordinary macroscopic system.

\section{Phase Space Analysis}

In this section we study the dynamical system behavior of the present model to understand 
it's asymptotic behavior\cite{stachowski}.  In order to form the
autonomous coupled differential equations, we define the dimensionless variables,
\begin{equation}\label{equ:41}
u=\frac{\rho_{m}}{3H^{2}}, \hspace{0.5in}
v=\frac{\rho_{de}}{3H^{2}}.
\end{equation}
 Using Friedmann equation (\ref{equ:4}) and conservation
equations (\ref{equ:5}) and (\ref{equ:6}), the autonomous differential equations can be obtained as,
\begin{figure}[pb]
 \centerline{\psfig{file=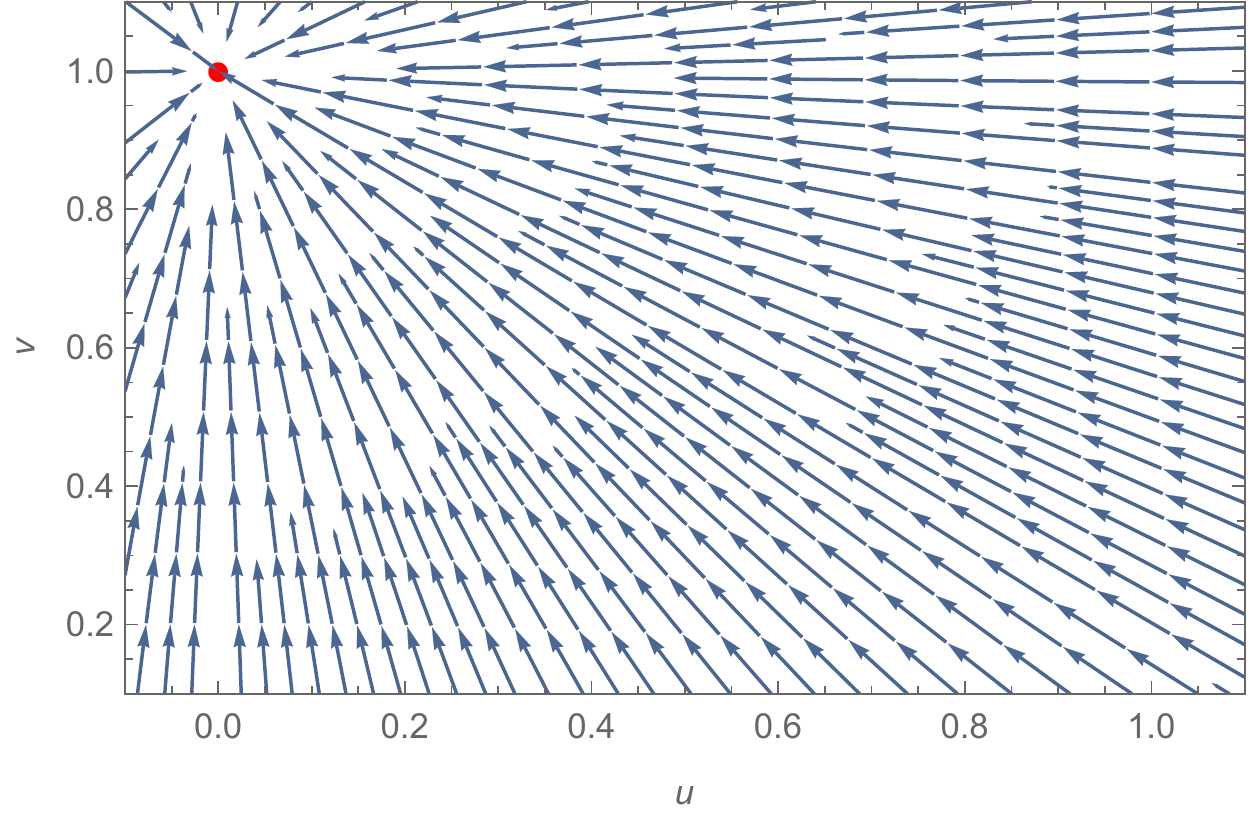,width=4.7cm}}
\vspace*{8pt}
\caption{The plot shows the phase space trajectory in the u-v plane
corresponding to the SNe+BAO+CMB data.The critical point in the
upper left corner of the plot is (0,1)is a future attractor .}\label{fig:f15}
\end{figure}
 \begin{equation}\label{equ:43}
u'=\frac{du}{dx}=3(b-1)u-6u(\frac{v}{2}-\frac{1}{2})=f(u,v),
\end{equation}
\begin{equation}\label{equ:44}
v'=\frac{dv}{dx}=-3bu-6v(\frac{v}{2}-\frac{1}{2})=g(u,v).
\end{equation}
Here the prime refers to a differentiation with the variable  $x=\ln a.$ The critical points,
 obtained by equating $u'=0$ and $v'=0,$ are
$(u_{\ast},v_{\ast})=(0,0),(1-b,b),(0,1),$  and are corresponding to the equilibrium solutions of the model. 
The first one, corresponding to the origin of the phase-space, refers to an empty universe or Milne universe\cite{milne}. 
The state (0,1) corresponding to  an end de Sitter epoch and point 
$(1-b,b)$ corresponds to a prior matter dominated era. 
The stability of these critical points is determined by the sign of the eigenvalue values of the Jacobian matrix, 
which is formulated by linearising the system of autonomous equations around each equilibrium points.

 Let us consider small perturbation about the critical points as, $u=u_{\ast}+\delta u$, $v=v_{\ast}+\delta v$. On linearizing the system of 
 equations (\ref{equ:43}) and (\ref{equ:44}) with respect to these perturbation, we get the matrix equation,
\begin{equation}\label{equ:45}
\begin{bmatrix}
\delta u'\\
\delta v'
\end{bmatrix}
=
\begin{bmatrix}
(\frac{\partial f}{\partial u})_{\ast} &(\frac{\partial f}{\partial
v})_{\ast}\\
(\frac{\partial g}{\partial u})_{\ast} &(\frac{\partial g}{\partial
v})_{\ast}
\end{bmatrix}
\begin{bmatrix}
\delta u\\
\delta v
\end{bmatrix}.
\end{equation}
 The $2\times 2$ matrix in the right side of
equation(\ref{equ:45}) is the Jacobian matrix and for the present model it takes the form,
\begin{equation}\label{equ:46}
\begin{bmatrix}
3(-1+b)-6(\frac{-1}{b}+\frac{v_{\ast}}{2}) &-3u_{\ast}\\
-3b &-6(\frac{-1}{2}+\frac{v_{\ast}}{2})-3v_{\ast}
\end{bmatrix}.
\end{equation}
Eigenvalues can be found by diagonalizing this matrix. 

\begin{table}[ph]
\renewcommand{\arraystretch}{2.3}
\caption{Critical points and eigenvalue values.}\centering
\begin{tabular}{|c |c |c| c| c| }
\hline
$Critical points$ & $Eigen values$ & $Critical points$ & $Eigen values$  & $Nature$ \\ 
  (WMAP7)         &    (WMAP7)     &   (Planck2013)    &     (Planck2013) &      \\ 
  \hline
(0,0) & (3,0.023) & (0,0) &  (3,0.027) & Unstable \\
(0.99,0.007)& (2.98,-0.022) & (0.99,0.009) &(2.97,-0.027) & Saddle \\
(0,1) &(-3,-2.98)& (0,1) & (-3,-2.973) & Stable \\ 
\hline
\end{tabular} \label{table:1}
\end{table}

A critical point is said to be stable, if the eigenvalues of the
Jacobian matrix evaluated at the critical point are all negative. In
such case the trajectories starting from the neighborhood of the critical point are all converges to it, irrespective of their initial conditions. Such points or solutions are called stable attractors.
If the eigenvalues are all positive,
then the critical point is unstable. Irrespective of the initial conditions, the trajectories emanating from the neighborhood of these points will diverge away. If the eigenvalues consists of 
positive and negative values, then the critical point is said to be saddle, for which the trajectories may converge or 
diverge depending up on the initial conditions\cite{Jamil}. 

The eigenvalues corresponding to the three critical points are obtained 
and are given in the Table \ref{table:1}. There are no differences in the characteristics of critical points obtained by the first data set and the second data set consisting of SNe, Planck 2013 and latest SDSS data.
It then follows that, critical point, (0,0) is unstable, 
since both the eigenvalue values of it 
are positive, while (0.99,0.007)(or (0.99,0.009) as per the second data set including Planck 2013 and latest SDSS data set) is a saddle point, since one of the eigenvalue values is positive and the other one is negative. The last point 
(0,1) is a future stable point, for which the eigenvalues are negatives according to both the data sets. The results are summarized in 
 Table \ref{table:1}. A phase space plot is shown figure (\ref{fig:f15}), which shows the convergence of all the trajectories into the future 
 attractor at (0,1).


\section{CONCLUSION}
Running vacuum models has been emerged as a strong alternative to cosmological constant models in explaining the recent acceleration of the universe. Earlier results indicating that generalized running vacuum energy can cause a transition from a prior deceleration to a late acceleration phase only with a nonzero constant in the vacuum energy density. 
In this work, by accounting the interaction between the running vacuum and the dark matter sector through the phenomenological term $Q=3bH\rho_m,$ we have shown that a transition from the prior decelerated to a late accelerated phase could be possible even without a constant additive term in the running vacuum density. Conventional running vacuum density consist of a combination of $\dot{H}$ and $H^2$ with two parameters. For simplicity we have considered the Ricci dark energy, also a combination of $\dot{H}$ and $H^2,$ with a single parameter, $\beta$ as a running vacuum.  We have analytically solved for the Hubble parameter and extracted the value of the interaction parameter $b$, model parameter $\beta$ and the present value of the Hubble parameter $H_0$ by constraining the model with the combined SNe+BAO+CMB observational data set.
The extracted values of the parameters are $\beta=0.46422^{+0.005}_{-0.004},$ $b=0.007615^{+0.01}_{-0.01}$, $H_0=70.06^{+0.53}_{-0.52}kms^{-1}Mpc^{-1}$ at the $1\sigma$ level. We repeat the computation by using SNe, Planck 2013 data and BAO data from the latest SDSS observations and obtained the parameters as $\beta \sim 0.4612,$ $b \sim 0.0094$ and $H_0 \sim 70.18kms^{-1}Mpc^{-1}.$ For comparison it is to be noted that by considering Ricci dark energy as the one with varying equation of state, the interaction parameter is found to be around $b=0.032^{+0.013}_{-0.013}$ at the $1\sigma$ level
in reference \cite{Fu1}.
But in reference\cite{Li3} the interaction parameter is evaluated as $b=- 0.00045\pm 0.00069$ at the $1\sigma$ level, by considering a general decaying cosmological vacuum energy. 

  The evolution of cosmological parameters like matter density and curvature scalar etc indicating the presence of
   big bang at the origin and an end de Sitter phase. The model predicts a transition from deceleration to acceleration epoch at the red shift $z_T=0.71$ and $z_T=0.74$ on using the second data set and is in good agreement with the observational result.
   
    The model is analyzed using the state finder diagnostic method. The evolution trajectory of the model in the state finder parameter plane, i.e. $(r,s)$ plane, is found to be restricted in the region corresponds to $r<1$ and $s>0$, which implies the quintessence nature of the running vacuum. The trajectory approaching the $\Lambda$CDM point in the end stages of the evolution. The present value of 
state finder parameters is found to be $(r_{0},s_{0})=(0.9907,0.00287)$ and $(r_{0},s_{0})=(0.9885,0.0035)$ on using second data set which are slightly different from the corresponding $\Lambda$CDM value $(1,0).$

The evolution of the horizon and matter entropy were studied, showing the huge increase in the horizon entropy is compensated by the decrease in the matter entropy, as a result the model obeys the generalized second law.
 
The dynamical system behavior of  the model has also been studied. Among the critical points obtained, one represents the 
prior matter dominated phase with decelerated expansion, with positive eigenvalue values hence is unstable. The second point, represents the later accelerating epoch, have negative eigenvalue values and hence is stable. The phase plot showing the convergence  of the trajectories in the neighborhood of the stable point is constructed. 

The low value of the interaction parameter indicating that the running of the vacuum with the Hubble parameter is comparatively slow. Such a slow decay of vacuum have speculated by many in the recent literature. Apart from the reference mentioned in the first paragraph in this section, another interesting study is in \cite{Salvatelli1}, where the authors have shown that the recent Plank data (along with SNe) favors a late time interaction between the dark sectors. They constrained the value of the interaction parameter, around 0.156, by using an interaction term, $Q=bH\rho_{\Lambda}.$ This value is slightly higher compared to our and in other references. But the previous authors have considered only a late interaction between the dark sectors.

\end{document}